# Development of Mental Models in Human-AI Collaboration: A Conceptual Framework

*Completed Research Paper*

**Joshua Holstein**
Karlsruhe Institute of Technology
Karlsruhe, Germany
Joshua.Holstein@kit.edu

**Gerhard Satzger**
Karlsruhe Institute of Technology
Karlsruhe, Germany
Gerhard.Satzger@kit.edu

## Abstract

*Artificial intelligence has become integral to organizational decision-making and while research has explored many facets of this human-AI collaboration, the focus has mainly been on designing the AI agent(s) and the way the collaboration is set up—generally assuming a human decision-maker to be "fixed". However, it has largely been neglected that decision-makers' mental models evolve through their continuous interaction with AI systems. This paper addresses this gap by conceptualizing how the design of human-AI collaboration influences the development of three complementary and interdependent mental models necessary for this collaboration. We develop an integrated socio-technical framework that identifies the mechanisms driving the mental model evolution: data contextualization, reasoning transparency, and performance feedback. Our work advances human-AI collaboration literature through three key contributions: introducing three distinct mental models (domain, information processing, complementarity-awareness); recognizing the dynamic nature of mental models; and establishing mechanisms that guide the purposeful design of effective human-AI collaboration.*

## Introduction

The rise of artificial intelligence (AI) has transformed the way organizations approach complex decisions, from healthcare (Prakash & Das, 2021; Yang et al., 2024) over finance (Vo et al., 2019) to manufacturing (Dengler et al., 2021). This transformation has evolved through distinct phases—from early rule-based systems (Liao, 2005) to today's advanced neural networks that can serve as collaborative partners with complementary strengths (Hemmer et al., 2025; Taudien et al., 2024). Unlike their predecessors, modern AI systems not only support human decision-making, but allow the formation of human-AI teams to collaboratively achieve shared goals (Van den Bosch et al., 2025). This evolution of AI systems presents two important effects with implications for human-AI collaboration:

First, AI systems introduce *additional challenges* for the human decision maker and their mental models, i.e., the "small scale models" (Craik, 1967) that individuals use to infer relationships, predict outcomes, and, more generally, understand the world around them (Johnson-Laird, 1983). Traditionally, decision-makers have relied on their mental models of the task domain to interpret information and, ultimately, make decisions (Chermack, 2003). System designers have created tools to augment this process through decision support systems that operated on explicit rule sets (Liao, 2005). As these systems essentially codified existing domain knowledge, their logic inherently aligned with users' existing mental models of the domain. However, the introduction of modern neural network-based AI systems with opaque decision logic (Abdel-Karim et al., 2023; Rai et al., 2019) requires decision-makers to develop supplementary mental models to understand how the AI systems process information and to recognize where human and AI capabilities complement each other to enable effective human-AI collaboration.

Second, AI systems do no longer merely augment human capabilities, but they also *reshape the cognitive processes* underlying human decision-making, i.e. how humans think about and solve problems (Lee et al., 2025, 2025). Specifically, they can trigger changes in human mental models. Despite the dynamic nature of mental models, research on human-AI collaboration has typically treated decision-makers as static recipients of AI recommendations. Discrepancies in decision performance were traced back to factors such as cognitive load (Schulz & Knierim, 2024), domain expertise (Bayer et al., 2022; Gnewuch et al., 2022), or cognitive styles (Rastogi et al., 2022; Riefle et al., 2024). Although researchers have begun exploring how humans can learn through interaction with AI systems (Förster et al., 2024a; Schemmer et al., 2023a; Spitzer et al., 2024b; van Zoelen et al., 2021) and how explanations can affect mental models (Bauer et al., 2023), these efforts remain limited and lack frameworks for understanding how AI system design can systematically influence the development of mental models.

Acknowledging this dynamic nature of mental models is crucial for effective human-AI collaboration. As mental models develop over time—triggered, for example, by the provision of contextual information (Spitzer et al., 2025) or by ongoing interactions with the AI system (Bauer et al., 2023)—it is imperative to understand how the characteristics of AI systems can shape this development (Van den Bosch et al., 2025). This centers on the question of how the purposeful design of AI systems can foster effective human-AI collaboration that improves mental models and preserves the complementary capabilities of the human decision-maker (Fügener et al., 2021; Hemmer et al., 2025), instead of diminishing it through continuous interaction. A key concern is that the decision-makers' mental models should not align with the AI system over time, producing overreliance and reducing the complementary potential in human-AI collaboration (Schemmer et al., 2022a). Instead, the design should promote the decision-maker's discretionary power to calibrate appropriate reliance on AI system recommendations (Schemmer et al., 2023c). To help close this gap in understanding and shaping human-AI collaboration, we pose the following research question:

**RQ:** *How can AI systems influence the development of mental models of decision-makers for effective human-AI collaboration?*

To address this research question, we follow a conceptual research approach (Gilson & Goldberg, 2015; Jaakkola, 2020; Whetten, 1989) to develop a socio-technical framework examining how AI systems shape decision-makers' mental models. Drawing on multiple theories from cognitive psychology (Endsley, 1995; Flavell, 1979; Johnson-Laird, 1983; Lim & Klein, 2006; Weick, 1995) and information systems (Leonardi, 2011), we identify three complementary and independent mental models relevant for human-AI collaboration—covering the *domain*, the AI system's *information processing* and the decision-maker's *complementarity-awareness*. Correspondingly, we propose three theoretically-grounded mechanisms that foster the development of these mental models. Based on these mechanisms, we develop testable propositions that hypothesize how each mechanism influences the development of its corresponding mental model, providing a theoretical foundation for future empirical research. Our framework, thus, makes a threefold contribution to human-AI collaboration literature: First, it proposes a perspective with three distinct mental models that systematically connect the understanding of the domain, the AI system's information processing, and complementarity-awareness. Second, it introduces a dynamic view of mental models of human decision makers—contrasting prevailing approaches that consider decision-makers as static recipients of AI recommendations. Third, it outlines mechanisms that drive the development of the proposed mental models, thus guiding the purposeful design of more effective human-AI systems.

The remainder of this paper is structured as follows: In Section 2, we first review the relevant literature on mental models and human-AI collaboration to establish the theoretical foundation for our framework. Next, we present our framework in Section 3, elaborating on how data contextualization, reasoning transparency, and performance feedback mechanisms influence mental model development before establishing propositions. We then discuss the implications of our framework for research and practice in Section 4, highlighting how it can guide the design of more effective human-AI collaboration, present a research agenda in Section 4, and conclude in Section 5.

## Foundations and Related Work

This section establishes the theoretical foundations for our framework on AI system influence on the mental model of decision-makers. We start with mental models as dynamic structures that guide decision-making and evolve through continuous interaction. We then explore human-AI collaboration in decision-making, before we specifically look at intersecting literature—covering mental models in collaborative human-AI decision-making.

## *Mental Models in Decision-Making*

Mental models represent simplified internal representations that individuals develop to understand and explain the world around them based on previous experiences (Johnson-Laird, 1983; Klein & Hoffman, 2008). These internal representations guide decision-making by directing attention toward relevant information while filtering out details perceived as extraneous (Endsley, 1995; Gary & Wood, 2005) to recognize familiar patterns that suggest appropriate solutions (Gavetti et al., 2005).

However, mental models are not static but continuously evolve through experience and learning (Tullio et al., 2007). When encountering new information, individuals filter it through their existing mental models to either facilitate integration of new knowledge or create resistance to contradictory evidence (Chermack, 2003; Harmon-Jones & Mills, 2019). Therefore, effective mental model development requires to resist this filtering tendency to allow the reconstruction of conceptual relationships (Argyris, 2017). This development process can deliberately be triggered by explicitly challenging existing beliefs (Kulesza et al., 2012; Pinski et al., 2023) or through continuous interaction (Critchfield & Twyman, 2014). Research consistently demonstrates that more sophisticated mental models, such as those possessed by domain experts compared to novices, directly correlate with superior decision performance through enhanced pattern recognition and problem-solving capabilities (Chi et al., 1981; Lurigio & Carroll, 1985).

A critical element in the effective application of mental models is metacognition—the awareness and understanding of one's own thought processes (Flavell, 1979). Metacognition enables decision-makers to monitor their cognitive processes, recognize limitations in their understanding, and adjust their approach accordingly (Yeung & Summerfield, 2012). Research has demonstrated that strong metacognitive abilities significantly enhance decision quality by allowing individuals to accurately assess confidence in their judgments (Ackerman & Thompson, 2017), to identify when additional information is needed (Hausmann & Läge, 2008), and to recognize situations where their existing mental models may be inadequate (Fleming, 2017; Fügener et al., 2022). This self-reflective capacity becomes especially valuable in complex decision environments where understanding the boundaries of one's expertise directly impacts decision quality.

Closely related to metacognition are the concepts of situation awareness and sensemaking, which explain how individuals process and interpret information in complex environments. Situation awareness theory explains how decision-makers process available information through multiple levels in real-time (Endsley, 1995). This involves recognizing critical information elements (Level 1) and understanding their relevance within the current decision context (Level 2) to infer future actions (level 3) (Endsley, 1995). Building on this immediate processing, sensemaking explains how individuals actively construct meaning from their experiences when confronted with novel or complex information that challenges their existing beliefs (Weick, 1995). It is the ongoing process through which people make retrospective sense of situations by extracting cues from their environment and using these as the basis for a plausible understanding that guides their actions. Together, these processes explain how individuals notice and bracket information, assess its relevance, and develop meaning that informs their decision-making (Brown et al., 2008; Endsley, 1995; Weick, 1995).

Beyond individuals, research has looked into mental models of teams (Mathieu et al., 2000; van den Bossche et al., 2011). Lim and Klein (2006) emphasize that effective collaboration requires team members to develop multiple complementary mental models: taskwork mental models that capture understanding of the task environment, and teamwork mental models that represent how capabilities and responsibilities are distributed across team members. These different mental models enable team members to anticipate needs of their teammates (Van den Bosch et al., 2025) and, ultimately, improve the team's decision performance (Mathieu et al., 2000; van den Bossche et al., 2011). Initial research has examined shared mental models in human-AI teams (Endsley, 2023; Schelble et al., 2022), looking how an understanding of the teammate's capabilities affects the collaboration (Van den Bosch et al., 2025). This consideration is especially relevant when examining decision-making in complex socio-technical systems where humans and AI systems work alongside each other (Kudina & van de Poel, 2024).

## Human-AI Collaboration in Decision-Making

Human-AI collaboration integrates humans and AI systems, forming a socio-technical system where effective outcomes depend on the dynamic interaction between them, rather than on isolated improvements in either one (Bansal et al., 2021; Kudina & van de Poel, 2024). Specifically, the effectiveness of human-AI collaboration emerges from the complementary strengths of humans and AI systems (Fügener et al., 2021; Hemmer et al., 2025): while humans possess contextual understanding and the ability of causal reasoning, AI systems are superior at detecting patterns in high-dimensional data and maintaining consistent decision criteria (Rastogi et al., 2023).

To leverage these complementary strengths, organizations have different options to implement human-AI collaboration that can generally be classified along two dimensions: first, the *decision method*, i.e. whether decision from humans and AI are integrated into an overall team decision or whether decision-making is delegated to just one team member; and second, the allocation of the (final) *decision authority*. Although all combinations across these dimensions are viable options, we specifically focus on approaches where humans retain the *ultimate decision authority*—as these configurations comply with recent regulatory frameworks such as the EU AI Act's requirement on human oversight (European Commission, 2023). This (integrated) decision requires the decision-maker to appropriately rely on the AI system, i.e., follow correct AI system advice, but trust one's own expertise when faced with flawed AI advice (Schemmer et al., 2023c). This calibration of *appropriate reliance* becomes particularly difficult when AI systems operate as "black boxes" with limited explainability (Rai et al., 2019). To address this challenge, transparency frameworks such as that proposed by (Lyons, 2013) suggest that humans need visibility into the system's analytical model, its capabilities and limitations to calibrate their reliance and effectively collaborate with the AI system overall.

## Mental Models in Collaborative Human-AI Decision-Making

Research on human-AI collaboration has increasingly recognized the importance of mental models, yet predominantly treats decision-makers as static recipients of AI recommendations rather than adaptive learners. We identify two main streams in existing literature:

The first stream establishes the *foundational importance of mental models* in human-AI collaboration. Bansal et al. (2019) demonstrate that accurate mental models of AI error boundaries significantly improve team performance, enabling humans to appropriately override AI recommendations. Similarly, Endsley (2023) emphasize the necessity of shared mental models between humans and AI systems, arguing that mutual understanding of capabilities is essential for effective collaboration. Together, these studies establish the importance of mental models in human-AI collaboration but typically conceptualize them as static constructs rather than studying their development over time.

The second research stream analyzes various *coordination mechanisms* governing collaborative human-AI decision-making. These mechanisms often implicitly impact mental models, while the mental model development is typically not explicitly addressed in the literature. Instead, a large body of research focuses on enhancing system understanding through explainable AI approaches (Förster et al., 2024a; Schemmer et al., 2022b; Spitzer et al., 2024b), on metacognition through confidence calibration (Ma et al., 2024; Taudien et al., 2024) and reflection (Abdel-Karim et al., 2023; Förster et al., 2024b; Jussupow et al., 2021), and on understanding the underlying data (Spitzer et al., 2025). While this body of research provides approaches for supporting decision-making, it typically evaluates effects on decision performance rather than examining how these mechanisms shape decision-makers' cognitive frameworks over time.

Only a limited subset of studies, though, explicitly examines the *development of mental models through interaction with AI systems*. Notably, Bauer et al. (2023) investigate how explanations can develop mental models, while Spitzer et al. (2024a) explore how incorrect explanations can distort decision-makers mental models. In recent work, van den Bosch et al. (2025) introduce a framework for co-learning in human-AI teams focusing on the design and effects of learning interventions in the context of urban search and rescue scenarios. Specifically, the authors highlight that co-learning requires the human and the AI to learn about the task being solved, about each other, and about task distribution across teammates. While they design specific learning interventions to learn about the capabilities of the partner such as the AI robot not being able to perform specific actions, they do not describe how the different learning areas can be systematically developed. Together, these studies represent important steps toward a dynamic understanding of mental model development. So far, however, they remain isolated efforts lacking a holistic and systematic understanding of the mechanisms that affect the development of mental models.

## Methodology

To address our research question on how AI systems influence decision-makers' mental models, we employ a conceptual research methodology. The primary goal of conceptual research is to integrate different streams of existing theory to develop an agreed-upon meaning about real-world phenomena by proposing new relationships among constructs through logical argumentation, rather than testing them empirically (Gilson & Goldberg, 2015; Whetten, 1989).

Our methodological approach follows established guidelines for conceptual theory development (Jaakkola, 2020) by systematically integrating previously disconnected research streams. We draw on mental model theory (Craik, 1967; Johnson-Laird, 1983) as our foundational framework for understanding how individuals develop internal representations of the world they interact with. To establish the three specific mental models relevant for human-AI collaboration, we utilize team mental model theory (Lim & Klein, 2006), which distinguishes between taskwork (domain) and teamwork (information processing and complementarity-awareness) mental models in collaborative contexts. To conceptualize how these mental models adapt and evolve through interaction with AI systems, we incorporate perspectives from sensemaking theory (Weick, 1995) and socio-technical systems theory (Leonardi, 2011), which views humans and technology as dynamically interrelated components rather than static entities.

For each of our three mental models we then propose one mechanism that drives their development. To do so, we draw on further theoretical foundations: situation awareness theory (Endsley, 1995) informs our data contextualization mechanism, explaining how decision-makers extract cues from their environment to construct domain understanding; Lyons' transparency framework (Lyons, 2013) grounds our reasoning transparency mechanism, establishing principles for making AI system operations understandable; and metacognition theory (Flavell, 1979) supports our performance feedback mechanism, explaining how external feedback calibrates complementarity-awareness. From these theoretical foundations, we derive testable propositions that hypothesize relationships between our identified mechanisms and mental model development outcomes, enabling future empirical validation (Jaakkola, 2020; Whetten, 1989).

## Conceptualizing Mental Models in Human-AI Collaboration

Building on our theoretical foundations, we now develop our framework that explains how decision-makers' mental models evolve through interaction with AI systems. Within this framework, we focus on AI systems that employ machine learning algorithms to learn patterns from data to generate predictions. This includes both traditional and interpretable approaches such as decision trees and linear models, as well as more recent complex techniques like neural networks and ensemble methods that often exhibit opacity due to their inherent complexity. We first identify the three mental models required for effective human-AI collaboration, before introducing the mechanisms through which AI systems influence these mental models and deriving testable propositions for future empirical research.

### *Existence of Complementary Mental Models*

Building on team mental model theory (Lim & Klein, 2006), we move beyond the prevailing conceptualization of mental models in human-AI collaboration research as a single, static construct and identify three complementary mental models essential for effective human-AI collaboration as shown in Figure 1.

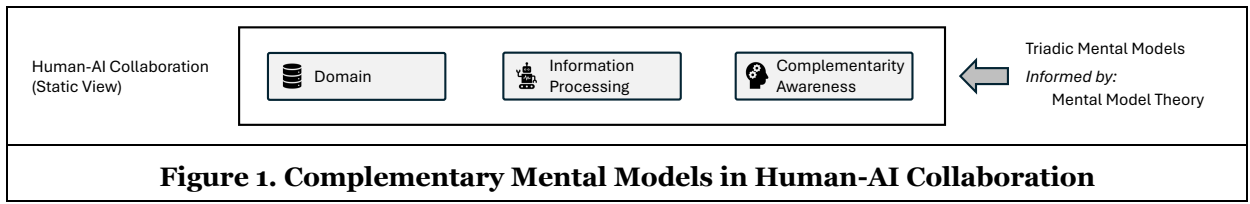

**Figure 1. Complementary Mental Models in Human-AI Collaboration**

**1) Domain mental models** represent the decision-maker's understanding of the underlying domain, including potential tasks and how collected data represents real-world phenomena (Chermack, 2003). Extending traditional taskwork mental models to the human-AI context, these domain models capture an understanding of relevant information for the task at hand (Endsley, 1995), relationships among attributes and target variables, the ability to distinguish meaningful patterns from noise, and the causal understanding that connects observed data to real-world phenomena. Understanding these inherent characteristics of the data allows decision-makers to contextualize and critically evaluate AI-generated recommendations against established domain knowledge.

**2) AI information processing mental models** capture how decision-makers understand the reasoning processes of the AI system providing recommendations (Endsley, 2023). Analogous to teamwork mental models but focused specifically on the AI system as a teammate, these models represent how AI systems transform inputs into recommendations and under what circumstances their outputs are reliable (Chromik et al., 2021; Endsley, 2023). Developing this mental model requires decision-makers to form an understanding of which inputs the AI system prioritizes, how it detects and interprets patterns, as well as the system's performance characteristics, including strengths, limitations, and potential biases across different cases (Chromik et al., 2021; Endsley, 2023).

**3) Complementarity awareness mental models** represent the decision-maker's understanding of how their own capabilities and limitations complement those of the AI system in the decision context (Ackerman & Thompson, 2017; Hausmann & Läge, 2008). Drawing on metacognition (Flavell, 1979), this involves assessing when one's expertise complements or exceeds the AI system's capabilities, when the AI system may be superior, and when to seek additional support (Risko & Gilbert, 2016). Complementing the teamwork mental model, decision-makers need to recognize not only their own potential biases and information processing limitations (Chermack, 2003), but understand these in relation to the AI system's strengths and weaknesses.

These complementary mental models connect with the fundamental components of human-AI collaboration: the data, the AI model, and the human decision-maker. Through this structural alignment, domain mental models primarily correspond with the data component, allowing decision-makers to interpret and contextualize inputs within their domain expertise and assess whether they appropriately capture relevant aspects of the task context. Information processing mental models correspond with the AI model, enabling decision-makers to develop expectations about how the system weighs different variables to generate its recommendations. Complementarity awareness mental models correspond with the decision-maker's own capabilities relative to the AI system, enabling metacognitive assessment of when their expertise complements or exceeds the AI system's capabilities, but also when the AI system may be superior. Together, these three mental models form a comprehensive framework to support understanding the complete collaborative human-AI system with its capabilities and limitations.

While we distinguish these three mental models analytically, they are interconnected in practice. Domain mental models may, for example, inform complementarity-awareness by revealing the boundaries of one's expertise within specific contexts, while simultaneously being refined through feedback about decision accuracy across different scenarios. Information processing mental models may shape domain understanding by highlighting which data patterns the AI system prioritizes, potentially revealing previously unnoticed relationships or challenging existing domain assumptions. Conversely, strong domain knowledge may enable more critical evaluation of AI system's information processing, supporting decision-makers in identifying when predictions conflict with established causal relationships. Complementarity awareness mental models may moderate how decision-makers integrate both domain knowledge and AI system understanding, determining when to trust personal expertise versus AI recommendations and when to seek additional information. Therefore, effective human-AI collaboration requires the simultaneous development across all three interdependent mental models rather than focusing on any single dimension. The following example illustrates how these interdependent mental models integrate in practice:

An engineer is charged to detect anomalies in a manufacturing process applying her domain mental model. Being assisted by an AI system flagging anomalies, her *mental model of information processing* recognizes that the AI system has identified an anomaly based on a sudden drop followed by a gradual rise in temperature. Her *domain mental model* recognizes that this most likely reflects windows having been opened in winter—an event unrelated to production quality in their context. Her *complementarity-* *awareness mental model*, thus, gives her the confidence to override the AI system. By integrating all three models, she accurately dismisses the alert without compromising quality control.

### *Dynamic Perspective of Mental Models*

As mental models continuously evolve when decision-makers encounter novel information that challenges their existing beliefs (Leonardi, 2011; Weick, 1995), mechanisms emerge that facilitate sensemaking processes by providing information for extraction, interpretation, and integration into existing mental frameworks (Brown et al., 2008; Weick, 1995): **data contextualization** shapes domain mental models (grounded in situation awareness theory; Endsley, 1995), **reasoning transparency** develops information processing mental models (grounded in transparency theory; Lyons, 2013), and **performance feedback** refines complementarity-awareness mental models (grounded in metacognition theory; Flavell, 1979). Figure 2 illustrates how these mechanisms foster the development of adaptive mental models in human-AI collaboration, providing a structured approach that both systematically influences mental model development and yields empirically testable propositions about how specific AI system design elements affect human-AI team performance. In the following, we describe the proposed mechanisms in detail.

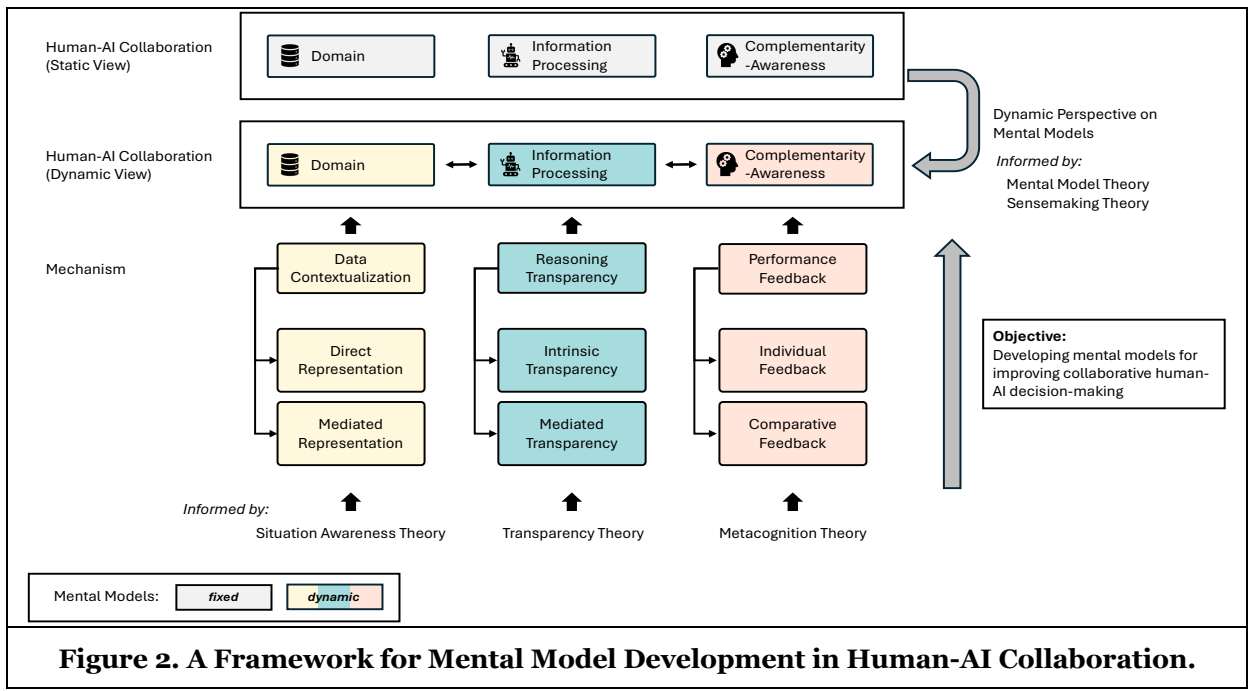

**Figure 2. A Framework for Mental Model Development in Human-AI Collaboration.**

## Data Contextualizing for Shaping Domain Understanding

Drawing on situation awareness theory (Endsley, 1995), we propose that domain mental models develop through mechanisms that provide context on underlying data patterns and relationships. Data contextualization mechanisms support both Level 1 and Level 2 situation awareness by highlighting relevant information elements and facilitating their understanding within the decision context (Endsley, 1995). To facilitate situation awareness, data contextualization informs decision-makers through two complementary approaches: direct and mediated representation of patterns and relationships.

**Direct Representation** focuses on presenting the data in ways that make inherent patterns and relationships more accessible. Through visualizations, statistical summaries, or interactive displays, this approach provides information such as statistical context (e.g., distributions or correlations) (Spitzer et al., 2025; Vössing, 2020) and structural information (e.g., hierarchies or relationships) (Kernan Freire et al., 2023).

**Mediated Representation** leverages algorithms to identify and highlight patterns in the data that might not be apparent through summaries or visualizations. This includes, for example, using techniques like clustering to identify groups with common characteristics (Arunachalam & Kumar, 2018). Further, AI systems can contextualize data through various types of explanations such as feature importance's (Nepal et al., 2024), counterfactuals (Schemmer et al., 2023b), example-based explanations (C. J. Cai et al., 2019), or uncertainty measures that highlight missing variables with explanatory power (Hüllermeier & Waegeman, 2021).

To ensure that data contextualization supports the development of domain mental models, it requires alignment with decision-makers' information needs, showing relevant relationships that inform their decision processes rather than providing generic information. While doing so, it needs to balance comprehensive representation against the decision-maker's limitations of information processing (Endsley, 1995, 2023), particularly in domains characterized by high-dimensional data such as manufacturing (Schemmer et al. 2023). In the human-AI context, this contextualizing process may enable decision-makers to develop a more accurate assessment of AI recommendations against domain knowledge. Accordingly, we formulate our first proposition:

**Proposition 1.** *Receiving data contextualization during decision-making allows decision-makers to develop their domain mental model by enhancing their understanding of relevant context, ultimately increasing human-AI team performance.*

## Reasoning Transparency for Shaping System Understanding

Building on Lyons's (2013) framework of transparency in human-robot interaction, we propose that mental models of AI information processing develop through mechanisms that make transparent how AI systems transform inputs into recommendations. This transparency requires decision-makers to understand the AI system's analytical model, i.e., its decision principles and inherent limitations (Lyons, 2013). Research in human-AI collaboration has shown that without this transparency, decision-makers often struggle to appropriately calibrate their reliance on AI outputs (Schemmer et al., 2023c; Spitzer et al., 2025). This challenge is amplified as modern AI systems are often complex and opaque (Lundberg & Lee, 2017), though the degree varies across different AI systems. To address this opacity, we propose transparency mechanisms that make AI systems more interpretable (Bayer et al., 2022) through two approaches: intrinsic transparency of inherently interpretable models and mediated transparency for opaque AI models.

**Intrinsic Transparency** utilizes inherently interpretable AI models such as decision trees, and linear models whose operations are directly observable (Adadi & Berrada, 2018). Decision-makers can therefore develop accurate mental models by directly observing how inputs translate into recommendations, for instance, by understanding conditional logic of decision trees or rule-based systems.

**Mediated Transparency**, conversely, employs techniques to make opaque "black-box" AI models more interpretable. While modern AI systems based on, amongst others, neural networks, transformers, or ensemble methods often achieve superior performance, they frequently sacrifice inherent transparency compared to simpler approaches (Adadi & Berrada, 2018; Lundberg & Lee, 2017). Therefore, they require explicit techniques to make their reasoning accessible. These can include explanation methods such as local interpretable model-agnostic explanations (LIME) (Ribeiro et al., 2016) or shapley additive explanations (SHAP) (Lundberg & Lee, 2017) that approximate model reasoning, interactive explanations that allow exploration through what-if scenarios (Bertrand et al., 2023), or uncertainty measures that highlight potential gaps in the coverage of the training data (Hüllermeier & Waegeman, 2021), i.e., the AI system's limitations (Lyons, 2013).

Both transparency approaches can support mental model development. While transparent models allow for direct observation and internalization of reasoning patterns, opaque models require decision-makers to construct mental models based on simplified representations of complex processes (Adadi & Berrada, 2018). By applying these transparency mechanisms—whether through inherently interpretable models or explanations of opaque ones—decision-makers can develop mental models that accurately represent how the AI system processes information (Bauer et al., 2023). This understanding enables better anticipation of AI system performance across contexts, allowing decision-makers to rely on AI recommendations when appropriate while overriding them when necessary. However, recent empirical evidence also reveals limitations of transparency interventions, for example, showing that explanations can have unintended consequences for decision fairness (Schöffer et al., 2024), with their effects depending critically on explanation design and implementation context, and can negatively impact decision quality and mental model development alike (Spitzer et al., 2024a). Therefore, we formulate our second proposition:

**Proposition 2.** *Receiving reasoning transparency during decision-making allows decision-makers to develop their information processing mental model by enhancing their understanding of how the AI transforms inputs into recommendations, ultimately increasing human-AI team performance.*

## Performance Feedback for Shaping Complementarity-Awareness Understanding

Drawing on metacognition theory (Flavell, 1979), we propose that performance feedback plays a critical role in developing accurate complementarity-awareness mental models. Since self-assessment without external feedback is often poorly aligned with actual performance (Harvey, 1997; Moore & Healy, 2008; Taudien et al., 2024), performance feedback can calibrate metacognitive awareness by helping individuals correct misunderstandings while building appropriate confidence in areas of strength (Butler et al., 2008). In human-AI contexts, metacognitive miscalibration can lead to inappropriate reliance decisions that fail to leverage the complementary strengths of human-AI teams (Ma et al., 2024; Taudien et al., 2024). This calibration enables decision-makers to understand complementary capabilities (Fügener et al., 2021; Hemmer et al., 2025) and develop appropriate reliance on the AI system's recommendations (Bansal et al., 2021; Schemmer et al., 2023c). We identify two complementary approaches to performance feedback: individual feedback on the own capabilities and contrastive feedback that juxtaposes human and AI capabilities across different decision contexts.

**Individual Feedback** provides decision-makers with objective information about own their performance patterns across different contexts, helping calibrate self-confidence to actual abilities (Ma et al., 2024). Effective individual feedback contextualizes outcomes within the domain, highlighting which problem types or data characteristics align with the decision-maker's knowledge, enabling recognition of domain-specific capabilities and limitations instead of presenting performance metrics alone.

**Comparative Feedback** contrasts human and AI performance to reveal complementary capabilities across decision contexts. Drawing on the complementarity strengths of humans and AI (Fügener et al., 2021; Hemmer et al., 2025; Rastogi et al., 2023), which suggests human-AI teams achieve superior performance when each agent contributes distinct strengths (Bansal et al., 2021; van den Bosch et al., 2025), this approach highlights scenarios where either human judgment or AI prediction systematically exceeds or fails across varying problem characteristics, for example through additional contextual knowledge of the human decision-maker.

Performance feedback requires careful design considerations, such as timing, framing, and specificity of feedback to support effective complementarity-awareness development (W. Cai et al., 2023; Miller et al., 2017; Sarwat Huma & Sarfraz, 2024). When carefully implemented, both individual and comparative feedback mechanisms may contribute to the development of complementarity-awareness mental models that enable decision-makers to understand when to rely on their own judgment versus AI recommendations, potentially enhancing the complementary integration of human and AI-systems. Therefore, we formulate our third proposition:

**Proposition 3.** *Receiving performance feedback during decision-making allows decision-makers to develop their complementarity-awareness mental model by enhancing their understanding of their own capabilities relative to the AI system, ultimately increasing human-AI team performance.*

### *Framework Synthesis*

Having established the three interconnected mental models and their development mechanisms, we now provide an integrated view of our framework in Table 1. It summarizes how each mental model develops through its mechanism grounded in established theory. However, as these mental models are interdependent, each mechanism's primary effects naturally also propagate among the interconnected mental models.

| Mental Model | Mechanism | Theoretical Foundation | Proposition |
|---|---|---|---|
| Domain | Data Contextualization | Situation Awareness Theory (Endsley, 1995) | Receiving data contextualization during decision-making allows decision-makers to develop their domain mental model by enhancing their understanding of relevant context, ultimately increasing human-AI team performance. |
| Information Processing | Reasoning Transparency | Transparency Theory (Lyons, 2013) | Receiving reasoning transparency during decision-making allows decision-makers to develop their information processing mental model by enhancing their understanding of how the AI transforms inputs into recommendations, ultimately increasing human-AI team performance. |
| Complementarity-Awareness | Performance Feedback | Metacognition Theory (Flavell, 1979) | Receiving performance feedback during decision-making allows decision-makers to develop their complementarity-awareness mental model by enhancing their understanding of their own capabilities relative to the AI system, ultimately increasing human-AI team performance. |

**Table 1. Synthesis of Mental Models, Mechanisms, and Propositions.**

# Discussion

This paper advances our understanding of human-AI collaboration by reconceptualizing decision-makers' mental models as adaptive components within socio-technical systems. At the core of our framework lies the recognition of three mental models that decision-makers develop when collaborating with AI systems: *domain mental models*, *information processing mental models*, and *complementarity-awareness mental models*. Unlike traditional approaches that view decision-makers as static recipients, we introduce how their mental models can be systematically shaped through specific mechanisms: domain mental models through data contextualization, information processing mental models through reasoning transparency, and complementarity-awareness mental models through performance feedback. We next position our framework within existing IS theory, examine its boundary conditions as well as outline a research agenda.

## *Theoretical Positioning and Boundary Conditions*

While our framework primarily focuses on how proposed mechanisms positively influence mental model development, we acknowledge that they may also have potential adverse effects by inducing behavior changes with negative impact on human-AI collaboration. Data contextualization that highlights misleading patterns might lead to flawed domain mental models as demonstrated by (Spitzer et al., 2024a), particularly when decision-makers lack the expertise to critically evaluate the information presented. Similarly, reasoning transparency mechanisms that oversimplify complex AI processes through explanations that do not accurately reflect the AI system's information processing may create distorted information processing mental models (Lakkaraju & Bastani, 2020), potentially leading to inappropriate reliance decisions. Finally, performance feedback mechanisms could amplify existing biases or create unwarranted confidence if feedback samples are limited or not representative of the broader decision space. These concerns are particularly relevant in the context of emerging AI systems like large language models, where hallucinations might systematically distort mental models in users who lack robust domain knowledge (Spitzer et al., 2024a).

Further, we acknowledge that additional factors may influence mental model development in human-AI collaboration. However, we argue that these three mechanisms are based on well-established cognitive theories and represent primary pathways through which decision-makers develop an understanding of the essential components of the socio-technical system during direct interaction with AI systems. While other AI design elements such as user control (Oz et al., 2023), or task complexity (Salimzadeh et al., 2023) may also influence mental model development, we argue that they often operate through these fundamental mechanisms, e.g., user interfaces present explanations or insights on the underlying data. Additional potential factors, such as social learning through peer observation (Rees et al., 2015) and formal training interventions (Pinski et al., 2023), may supplement these primary mechanisms. However, these external factors still influence mental model development by shaping how decision-makers engage with and interpret the information provided through our primary mechanisms. For example, training might enhance a decision-maker's ability to interpret data contextualization, while peer learning might support decision-makers in their assessment how to leverage the complementary strength of themselves and the AI system by better understanding the AI system's information processing. By focusing on our presented mechanisms, our framework provides mechanisms that can be directly instantiated in AI systems to systematically develop decision-makers' mental models.

Additionally, our framework treats decision-makers as a relatively homogeneous group, yet extensive research in human-AI collaboration has identified various moderating factors that influence how people interact with AI systems. These include, amongst others, domain expertise levels that shape how individuals interpret AI advice and explanations (Morrison et al., 2024), trust propensity that influences initial attitudes toward AI recommendations (Schemmer et al., 2023c), and cognitive styles that determine preferences for different types of information presentation and reasoning approaches (Riefle et al., 2024). Furthermore, individuals vary in their responsiveness to different intervention strategies, such as confidence scores and uncertainty information (Holstein et al., 2025) or specific explanation formats such as feature-based, example-based, or natural language descriptions (Schöffer et al. 2024; Buçinca et al. 2025). While our framework proposes general mechanisms for mental model development, the effectiveness of data contextualization, reasoning transparency, and performance feedback likely varies considerably across individuals based on these established moderating factors. Future empirical validation should explicitly account for such individual differences to understand boundary conditions and improve mechanism design for diverse user populations.

Our emphasis on mechanisms integrated within AI systems aligns with our dynamic view of mental models through interaction with these systems, which challenges the prevailing assumption in human-AI collaboration research that treats both decision-makers and AI systems as fixed entities. Current research predominantly focuses on understanding various human factors (e.g. Bayer et al., 2022; de Zoeten et al., 2024; Gnewuch et al., 2022; Rastogi et al., 2022; Riefle et al., 2024; Schemmer et al., 2023c; Schulz & Knierim, 2024), with only little research focusing on the influence of the AI-systems on human mental models (e.g., Abdel-Karim et al., 2023; Bauer et al., 2023; Ma et al., 2024; Spitzer et al., 2024a; van den Bosch et al., 2025). We address this limitation by examining the essential mental models' decisions-makers develop through continuous interaction with AI systems and theorize how these mental models can be systematically shaped through the purposeful design of AI systems.

Building on this dynamic perspective of mental models, our framework aligns with (Leonardi, 2011) concept of imbrication—the interweaving of human and material agencies. Through this lens, decision-makers develop their mental models through interaction with the AI system, consequently changing their decision-making routines: they may approach problems differently or refine their information needs based on their evolving understanding. These changing routines subsequently create demand for modifications to AI systems, such as incorporating new training data or developing novel types of explanations. While our research focuses specifically on how AI systems influence the mental models that decision-makers develop, it establishes a foundation for understanding the full imbrication cycle where evolving mental models may ultimately drive changes in AI system design. This approach extends beyond immediate decision-making performance to consider the long-term effects of AI systems on human mental model development, and, in the long term, also the evolution of the socio-technical system as a whole.

Finally, the interdependence of the three introduced mental models reveals a potential reason why interventions focusing exclusively on one dimension (e.g., explanations without domain contextualization or domain expertise without an adequate explanation of how the AI systems generate their recommendations) often fail to realize the full potential of human-AI collaboration (Hemmer et al., 2021). Insufficiently developed domain mental models can leave users unable to effectively verify AI recommendations, for example, when decision-makers are unable to identify hallucinations of large language models like GPT (Wang et al., 2024). Similarly, inadequate information processing mental models hinder the understanding of the AI system's reasoning process, potentially resulting in either underreliance or even resistance to AI system adoption (Bayer et al., 2022). Finally, underdeveloped complementarity-awareness mental models may result in users failing to leverage their unique strengths that could complement the AI system (Ma et al., 2024), leading to suboptimal human-AI collaboration outcomes.

## *Research Agenda*

**Empirically Validating Framework Propositions.** Our conceptual work hypothesizes concepts and interrelations for mental model development. It provides the base for empirical validation to test our propositions – and confirm the influence of the mechanisms on mental model development and, ultimately, also on the human-AI team performance. Longitudinal studies can help to understand how these mechanisms shape mental models over time in continuous interaction with AI systems. While our framework suggests that decision-makers continuously refine their understanding through interaction with AI systems, we lack empirical evidence of how our proposed mechanisms interact over time and how they differ in their effects on immediate performance and long-term mental model development.

**Covering Different Types of Mental Models.** Mapping our framework against the existing literature, we find that coverage of different mental models is very unbalanced in the literature: While explanations supporting information processing models have received substantial scholarly attention (Schemmer et al., 2022a), mechanisms for domain model development through data contextualization remain largely unexplored. Similarly, approaches for developing complementarity-awareness models through performance feedback have only recently begun to attract research interest (Abdel-Karim et al., 2023; Ma et al., 2024; Taudien et al., 2024). These understudied mechanisms, therefore, require systematic investigation regarding their influence on mental model development and human-AI team performance, including potential adverse effects, for example, from incorrect AI outputs, misleading explanations or hallucinations that might systematically distort mental models.

**Investigating Cross-Mechanism Effects on Mental Model Development.** While we acknowledged earlier that mental models are interdependent, our framework does not explore whether mechanisms might also have direct cross-effects beyond their primary targets. For example, performance feedback might directly influence domain understanding by revealing accuracy patterns across different domain contexts, rather than only working through complementarity-awareness mental models. Future research should empirically investigate these direct cross-mechanism pathways to capture the complexity of interactions in human-AI collaboration, strengthening the socio-technical systems perspective that underpins our work.

**Investigating Reciprocal Relationship of Human and AI System Evolution.** Our framework emphasizes how AI systems shape human mental models but does not fully address the reciprocal relationship suggested by the imbrication perspective (Leonardi, 2011). While we acknowledge that evolving mental models may drive changes in AI system design, we do not explore the mechanisms through which decision-makers' developing understanding might influence subsequent iterations of the AI system, i.e., how human decision-makers and AI adapt to each other over time (Holstein et al., 2020; van den Bosch et al., 2025). Future research should examine how changes in mental models lead to modified data collection practices, different interpretations of AI outputs, or demands for new system capabilities. This bidirectional relationship between human cognitive development and AI system refinements represents a promising area for future research that could advance our understanding of the co-evolution of human-AI socio-technical systems (Berente et al., 2021; Kudina & van de Poel, 2024), where also AI systems understand when human are more likely to be correct (Endsley, 2023).

**Extending Framework to Other Human-AI Collaboration Configurations.** Our framework primarily addresses approaches where humans maintain decision authority and AI primarily provides recommendations. This focus does not capture the full spectrum of human-AI configurations, particularly those where AI systems might obtain greater agency, e.g., when AI systems can delegate instances to the human decision-maker (Fügener et al., 2022; Westphal et al., 2025). As AI capabilities advance, future research should examine whether mental model development differs across various human-AI interaction paradigms, including those where AI systems may possess decision authority.

**Incorporating Organizational Context Factors.** The framework so far does not account for the organizational context in which human-AI collaboration occurs. Previous research has shown that organizational factors such as incentive structures (Roumani et al., 2015), hierarchical dynamics (Chelmis & Prasanna, 2013), and organizational cultures (Reis et al., 2020) can influence the adoption of technology and AI systems and, consequently, also the development of appropriate mental models. Future research should examine how these organizational factors moderate the relationship between our identified mechanisms and mental model development. Beyond examining organizational factors, future research should also investigate whether mental model development processes differ in collaborative contexts compared to individual settings. This includes exploring how the configuration of human-AI interaction changes these processes, such as when one AI system is used by teams rather than individual collaborations of one human with an AI system. In collaborative environments, decision-makers may need to develop both individual mental models of the AI system and shared mental models with teammates, potentially altering how data contextualization, reasoning transparency, and performance feedback influence mental model development through collective sensemaking processes.

# Conclusion

To summarize, this paper addresses our research question by proposing a framework that introduces three distinct mental models—domain, information processing, and complementarity-awareness—that form the foundation for effective human-AI collaboration, each developed through specific mechanisms: *data contextualization*, *reasoning transparency*, and *performance feedback*. Moving beyond static perspectives that treat humans as mere recipients of AI recommendations, we establish a dynamic socio-technical view where decision-makers' mental models are shaped through interaction with the AI system. This perspective extends how we evaluate AI systems beyond accuracy metrics to include their capacity to enrich decision-makers' understanding of both the domain and the AI's capabilities. Future research should empirically validate our propositions, testing both immediate task performance improvements and long-term effects on mental models. As organizations increasingly deploy AI for complex decisions, this extended perspective offers a more comprehensive approach to designing complementary socio-technical systems that leverage the strengths of both human and artificial intelligence.

*Mental Models in Human-AI Collaboration: A Conceptual Framework*